\documentclass[sigconf,natbib=true]{acmart}

\AtBeginDocument{%
}

\usepackage{enumitem}
\usepackage{xspace}
\usepackage{subcaption}
\usepackage{multirow}
\usepackage{colortbl}
\usepackage{array}
\usepackage{booktabs}
\newcolumntype{C}[1]{>{\centering\arraybackslash}p{#1}}
\usepackage{xspace}
\usepackage{xcolor}
\usepackage{soul}
\usepackage{pifont}
\usepackage{dsfont}
\usepackage{tabularx}
\usepackage[many]{tcolorbox}

\begin{document}

\title[KGMEL: Knowledge Graph-Enhanced Multimodal Entity Linking]{KGMEL: Knowledge Graph-Enhanced Multimodal Entity Linking}

\settopmatter{authorsperrow=4}
\author{Juyeon Kim}
    \affiliation{%
        \institution{KAIST}
        \city{Seoul}
        \country{Republic of Korea}
    }
    \email{juyeonkim@kaist.ac.kr}

\author{Geon Lee}
    \affiliation{%
	\institution{KAIST}
        \city{Seoul}
        \country{Republic of Korea}
    }
    \email{geonlee0325@kaist.ac.kr}

\author{Taeuk Kim}
    \authornote{Co-corresponding authors.} 
    \affiliation{%
	\institution{Hanyang University}
        \city{Seoul}
        \country{Republic of Korea}
    }
    \email{kimtaeuk@hanyang.ac.kr}
	
\author{Kijung Shin}   
    \authornotemark[1]
    \affiliation{%
	\institution{KAIST}
        \city{Seoul}
        \country{Republic of Korea}
    }
\email{kijungs@kaist.ac.kr}



\copyrightyear{2025}
\acmYear{2025}
\setcopyright{cc}
\setcctype{by}
\acmConference[SIGIR '25]{Proceedings of the 48th International ACM SIGIR Conference on Research and Development in Information Retrieval}{July 13--18, 2025}{Padua, Italy}
\acmBooktitle{Proceedings of the 48th International ACM SIGIR Conference on Research and Development in Information Retrieval (SIGIR '25), July 13--18, 2025, Padua, Italy}\acmDOI{10.1145/3726302.3730217}
\acmISBN{979-8-4007-1592-1/2025/07}

\begin{abstract}
Entity linking (EL) aligns textual mentions with their corresponding entities in a knowledge base, facilitating various applications such as semantic search and question answering.
Recent advances in multimodal entity linking (MEL) have shown that combining text and images can reduce ambiguity and improve alignment accuracy.
However, most existing MEL methods overlook the rich structural information available in the form of knowledge-graph (KG) triples.

In this paper, we propose \method, a novel framework that leverages KG triples to enhance MEL.
Specifically, it operates in three stages:
\textbf{(1) Generation:} Produces high-quality triples for each mention by employing vision-language models based on its text and images.
\textbf{(2) Retrieval:} Learns joint mention-entity representations, via contrastive learning, that integrate text, images, and (generated or KG) triples to retrieve candidate entities for each mention.
\textbf{(3) Reranking:} Refines the KG triples of the candidate entities and employs large language models to identify the best-matching entity for the mention.
Extensive experiments on benchmark datasets demonstrate that \method outperforms existing methods. Our code and datasets are available at: \url{https://github.com/juyeonnn/KGMEL}.

\end{abstract}

\newcommand{\smallsection}[1]{\vspace{0.2pt}{\noindent {\bf{\underline{\smash{#1}}}}}}

\newcommand{\ours}{\textsc{KGMEL}\xspace}
\newcommand{\method}{\textsc{KGMEL}\xspace}
\newcommand{\cmt}[1]{\textcolor{red}{#1}}
\newcommand{\bmt}[1]{\textcolor{blue}{#1}}
\newcommand{\xmark}{\ding{55}}

\newcommand\red[1]{\textcolor{red}{#1}}
\newcommand\blue[1]{\textcolor{blue}{#1}}

\setlength{\textfloatsep}{0.12cm}
\setlength{\dbltextfloatsep}{0.12cm}
\setlength{\abovecaptionskip}{0.12cm}
\setlength{\skip\footins}{0.12cm}

\newtcolorbox{prompt}{
    boxsep=1pt, 
    boxrule = 1pt,
    rounded corners,
    arc = 2pt  
}
\maketitle

\section{Introduction \& Related Work}
\label{sec:intro}
Entity linking (EL) aims at aligning \textit{mentions} (phrases within a document) with their corresponding \textit{entities} in a knowledge base, supporting various applications, including semantic search~\cite{cheng2007entityrank,meij2014entity,bordino2013penguins,liao2021mmconv}, question answering~\cite{xiong2019improving,longpre2021entity}, and dialogue systems~\cite{cui2022openel,curry2018alana}.

\begin{figure}[t]
    \centering

    \includegraphics[width=0.48\textwidth]{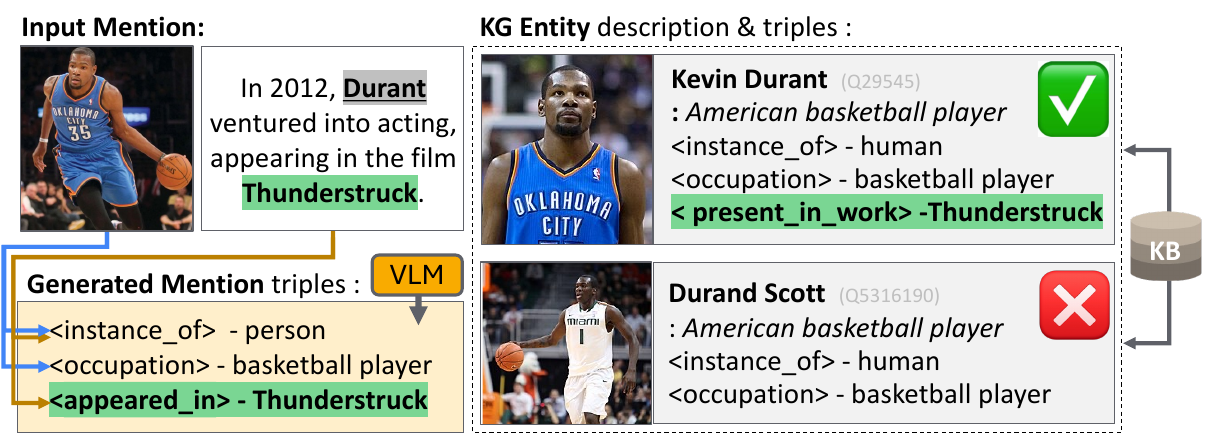} 
        \caption{An example of multimodal entity linking (MEL) using \ours. \ours generates triples for the mention to be matched with knowledge graph (KG) triples in the knowledge base (KB). In the figure, blue and yellow arrows point to triples derived from visual and textual context, respectively. 
        }
    \label{fig:case}
\end{figure}

Recently, multimodal entity linking (MEL)~\cite{moon2018multimodal} has emerged, integrating textual and visual information from mentions and entities to reduce ambiguity and improve linking accuracy.
Many approaches focus on learning representations of text and images in a shared latent space
~\citep{adjali2020multimodal,wang2022multimodal,luo2023multi,hu2024multi}. 
 Other methods leverage the in-context capabilities of large language models (LLMs) to perform zero-shot or few-shot matching 
 \citep{shi2023generative,long2024trust,wang2023benchmarking,liu2024unimel}.

However, knowledge-graph (KG) triples, which provide a structured representation of an entity, remain largely overlooked in MEL. 
We observe that \textbf{(O1)} entities in common knowledge bases typically have more and longer triples than textual descriptions, and \textbf{(O2)} triples offer essential context for disambiguating otherwise indistinguishable entities (see Section~\ref{sec:prelim}). These observations motivate our exploration into integrating KG triples into MEL. 

Incorporating KG triples into MEL, however, presents several challenges.
\textit{First}, mentions do not inherently possess associated triples, making a direct comparison with KG triples non-trivial.
\textit{Second}, while each entity in the knowledge base is associated with a large number of triples, only a small subset of them is relevant for linking, while the rest may be redundant or even noisy. 

To address these challenges, we propose \method, a novel framework that leverages KG triples to enhance MEL, as illustrated in Figure~\ref{fig:case}.
Specifically, \method consists of three stages:

\begin{itemize}[leftmargin=*]
    \item \textbf{(Stage 1) Generation.} 
    For each mention, \method generates high-quality triples from the mention's textual and visual information, using vision-language models (VLMs).
    \item \textbf{(Stage 2) Retrieval.} 
    \method learns joint representations of mentions and entities integrating text, images, and (generated or KG) triples, optimized via contrastive learning to align relevant entities with the mention.
    Then, it uses these embeddings to retrieve a subset of candidate entities in the knowledge base.
    \item \textbf{(Stage 3) Reranking.} 
    \method refines each candidate by filtering out irrelevant KG triples, retaining only those most pertinent to the mention.
    Then, it determines the best-matching entity based on the filtered triple information, leveraging LLMs.
\end{itemize}

\noindent
Our experimental results demonstrate that \method outperforms the state-of-the-art MEL baselines in three benchmark datasets.

Our contributions are summarized as follows:
\begin{itemize}[leftmargin=*]
    \item \textbf{Observations.} We quantitatively and qualitatively analyze KG triples in real-world knowledge bases and their potential for MEL.
    \item \textbf{Method.} We propose \method, a novel generate-retrieve-rerank framework that effectively leverages triples for MEL.
    \item \textbf{Experiments.} \method outperforms the best competitor by up to 19.13\% in terms of HITS@1.
\end{itemize}

\section{Problem Definition \& Data Analysis}
\label{sec:prelim}
\subsection{Problem Definition}
Given a set of entities $\mathcal{E}$ in a multimodal knowledge base, each entity $e\in \mathcal{E}$ is represented as $\{t_e, v_e, \mathcal{T}_e\}$, where $t_e$ denotes the textual context, $v_e$ denotes the visual context, and $\mathcal{T}_e$ is the set of knowledge-graph (KG) triples in which $e$ is the head. 
A mention $m$ is represented as $\{t_m,v_m\}$, where $t_m$ and $v_m$ are its textual and visual information, respectively. 
Note that the mention does not include any triple information.
Given a mention $m$,
the goal of MEL is to identify the ground-truth entity $e_m\in \mathcal{E}$ that best matches $m$. 

\begin{figure}[t]
    \centering
    \hspace{-2mm}
    \includegraphics[width=0.49\textwidth]{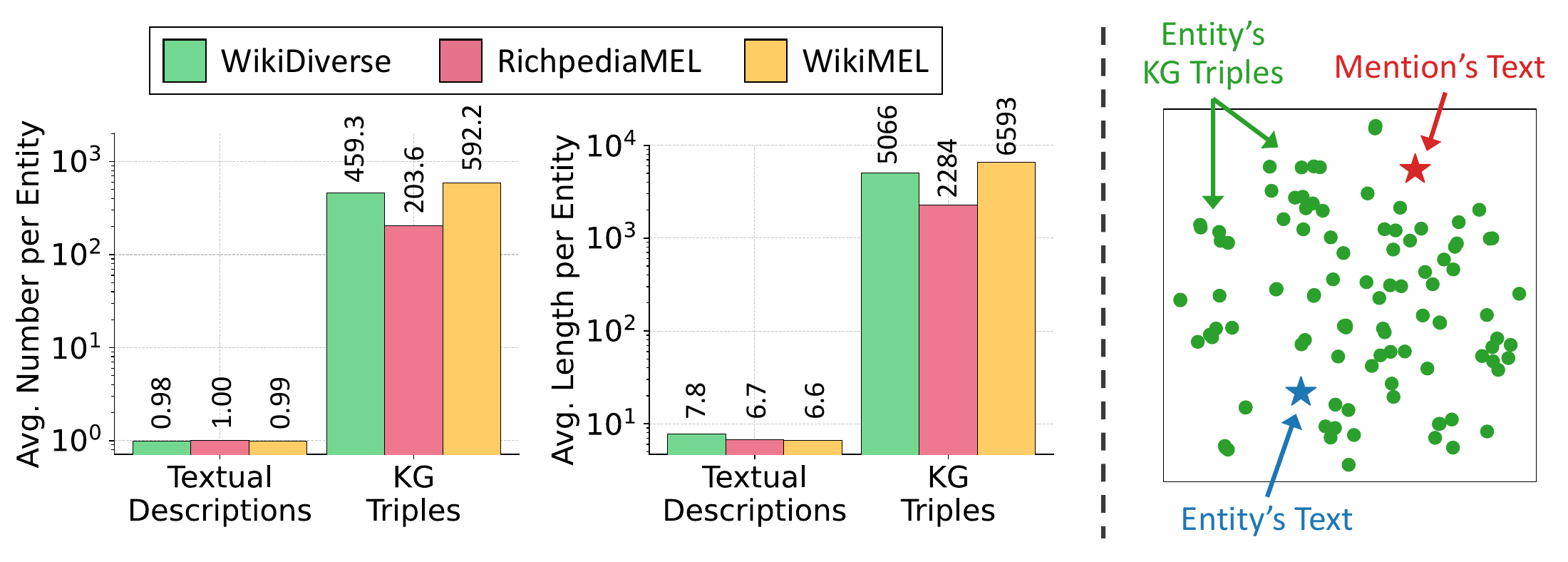}
    \caption{(Left) Comparison of the average number and word length of descriptions and triples per entity across WikiDiverse, RichpediaMEL, and WikiMEL datasets. (Right) t-SNE visualization illustrating the contextual similarity between mention sentences, entity descriptions, and entity triples.}
    \label{fig:triple-anlysis}
\end{figure}

\subsection{Data Analysis: Triples in Knowledge Bases}
We analyze multimodal knowledge bases, WikiDiverse~\citep{wang2022wikidiverse}, RichpediaMEL~\citep{wang2022multimodal}, and WikiMEL~\citep{wang2022multimodal}, presenting two key observations.

\smallsection{(O1) Abundance of KG triples.}
Knowledge bases contain a vast number of KG triples. 
As shown in Figure~\ref{fig:triple-anlysis}, each entity typically has a single concise textual description, while the number of associated triples averages in the hundreds. 
Moreover, the total length of these triples is substantially greater than that of textual descriptions, indicating their potential as a rich source of entity information. 

\smallsection{(O2) Triples as semantic bridges.}
KG triples provide contextual information that links mentions to entities that would otherwise remain unmatched when relying only on textual descriptions.
In Figure~\ref{fig:triple-anlysis}, we visualize embeddings of a mention's text, its corresponding entity's text, and its associated triples, obtained using a pretrained BERT model~\cite{devlin2018bert} and projected using t-SNE~\cite{van2008visualizing}.
While the mention and entity text embeddings are distant in the latent space, triple embeddings can be used as a \textit{semantic bridge} to bring them closer together.
This demonstrates how triples complement textual descriptions by capturing additional semantic information.

\section{Proposed Method: \method}
\label{sec:method}
\begin{figure*}[t]
    \vspace{-3mm}
    \centering
    \includegraphics[width=
    \textwidth]{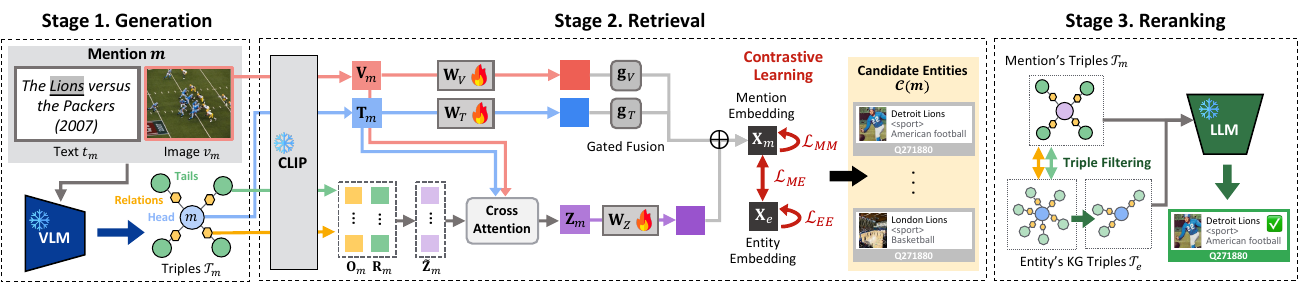} 
    \caption{Overview of \method. Our framework consists of three stages: \underline{\smash{(1) Generation:}} We generate triples for mentions using VLMs. \underline{\smash{(2) Retrieval:}} We obtain joint embeddings by integrating textual, visual, and triple-based embeddings, and using them, we retrieve $K$ candidates. \underline{\smash{(3) Reranking:}} 
    After filtering out irrelevant KG triples and retaining only those relevant to the mention, for each candidate, we determine the best-matching entity using LLMs.}
    \label{fig:pipeline}
\end{figure*}

In this section, we present \method (Figure~\ref{fig:pipeline}), a generate-retrieval-and-rerank framework that leverages triples to enhance MEL.


\subsection{Stage 1. Triple Generation of Mentions}
\label{sec:kg-gen}
Since mentions lack any associated triples, we generate triples for the mention by leveraging both textual and visual descriptions. 
However, extracting meaningful triples that effectively capture the mention's details is non-trivial due to the complexity of integrating and interpreting information across the textual and visual modalities.
To this end, we employ vision-language models (VLMs), which are trained on large-scale multimodal datasets and thus exhibit strong zero-shot reasoning and understanding capabilities, without requiring extensive retraining or fine-tuning.

Formally, given a mention $m$ with multimodal information $\{t_m,v_m\}$, we generate the set of triples ${\mathcal{T}}_m$ as follows:
\begin{equation*}
    {\mathcal{T}}_m = \textbf{VLM}\left(P_{\text{triple}}(t_m, v_m)\right),
\end{equation*}
where $P_\text{triple}$ is the prompt that instructs the VLM to generate triples with $m$ as the head.
Specifically, we design $P_\text{triple}$ to prompt the VLM step-by-step: 
(1) identifying the mention's type according to named entity recognition (NER), 
(2) describing the mention, and 
(3) generating the triples that provide a structured representation of the mention. 
Furthermore, we observe that providing VLMs with some representative relation examples during triple generation helps produce more comprehensive and accurate triples. 
For more details on the prompting strategy, refer to Appendix \ref{app:triple-gen}.

\subsection{Stage 2. Candidate Entity Retrieval}
\label{sec:entity-retrieval}
Now that we have complete data (spec., text, images, and triples) for both mentions and entities, we learn their representations by jointly leveraging all three data. 
The resulting representations for mentions and entities are then used to retrieve candidate entities for any given mention. 
Below, we describe how \method encodes mentions, while the same approach applies when encoding entities. 

\smallsection{Text \& image encoding.}
Consider a mention $m$ with associated textual, visual, and generated triple information $\{t_m,v_m,\mathcal{T}_m\}$.
We first encode its textual component $t_m$~\footnote{In practice, we further enhance the mention's text using VLMs.} and the visual component $v_m$ using a pretrained CLIP model~\cite{radford2021learning}.
Specifically, we obtain the text embedding $\mathbf{T}_m$ and the image embedding $\mathbf{V}_m$ as follows:
\begin{equation*}
    \mathbf{T}_m=\textbf{CLIP}(t_m)\in \mathbb{R}^{d'}\;\;\;\text{and}\;\;\;\mathbf{V}_m=\textbf{CLIP}(v_m)\in \mathbb{R}^{d'},
\end{equation*}
where we use the [\texttt{CLS}] token embeddings for both modalities.

\smallsection{Triple encoding.}
For the set of triples $\mathcal{T}_m$, we encode its relations and tails into embedding matrices $\mathbf{R}_m \in \mathbb{R}^{|\mathcal{T}_m|\times d'}$ and $\mathbf{O}_m \in \mathbb{R}^{|\mathcal{T}_m|\times d'}$, respectively.
Concretely, for the $i$\textsuperscript{th} triple $(m,r_i,o_i)\in \mathcal{T}_m$, the $i$\textsuperscript{th} rows of $\mathbf{R}_m$ and $\mathbf{O}_m$ are obtained via CLIP, i.e., $\mathbf{R}_{m,i} = \textbf{CLIP}(r_i)\in \mathbb{R}^{d'}$ and $\mathbf{O}_{m,i} = \textbf{CLIP}(o_i)\in \mathbb{R}^{d'}$.
We then combine these embeddings to construct the triple embedding matrix $\widetilde{\mathbf{Z}}_m$ as:
\begin{equation*}
    \widetilde{\mathbf{Z}}_m = \mathbf{O}_m + \textbf{MLP}([\mathbf{O}_m \;||\; \mathbf{R}_m]) \in \mathbb{R}^{|\mathcal{T}_m|\times d'},
\end{equation*}
where $||$ denotes concatenation.
The residual connection preserves tail information $\mathbf{O}_m$ while enriching it with relational context $\mathbf{R}_m$.

Having obtained $\widetilde{\mathbf{Z}}_m$, which contains embeddings for the $|\mathcal{T}_m|$ triples, we aggregate them into a single representative triple embedding for $m$.
Since triples vary in importance and their significance depends on other modalities (i.e., text and images), we compute a dual cross-attention score $s_m$ as follows:
\begin{equation*}
    s_m = \textbf{Softmax}\left( \frac{\beta\; \widetilde{\mathbf{Z}}_m \mathbf{T}_m^\top + (1-\beta)\;  \widetilde{\mathbf{Z}}_m \mathbf{V}_m^\top }{\tau_\text{att}} \right)\in (0,1)^{|\mathcal{T}_m|},
\end{equation*}
where $\beta$ is a hyperparameter that balances the contributions of the two modalities, and $\tau_\text{att}$ is a temperature term. 
The resulting scores $s_m$ capture the relevance of each triple relative to both textual ($\mathbf{T}_m$) and visual ($\mathbf{V}_m$) modalities.
To further denoise the attention scores, we retain only the top-$p$ values in $s_m$ to form $\hat{s}_m$, i.e., $\hat{s}_{m,i} = s_{m,i}\cdot\mathds{1}[s_{m,i}\;\text{is in the top-$p$ of $s_m$}]$ where $\mathds{1}[\cdot]$ is an indicator function.
Finally, we aggregate the triple embeddings into a single vector as:
\begin{equation*}
    \mathbf{Z}_m = \sum\nolimits_{i=1}^{|\mathcal{T}_m|} \hat{s}_{m,i} \;\widetilde{\mathbf{Z}}_{m,i} \in \mathbb{R}^{d'}.
\end{equation*}
By applying the same mechanism to each entity $e\in \mathcal{E}$, we can similarly obtain $\mathbf{T}_e$, $\mathbf{V}_e$, and $\mathbf{Z}_e$.

\smallsection{Gated fusion.}
Given the textual ($\mathbf{T}_m$), visual ($\mathbf{V}_m$), and triple-based ($\mathbf{Z}_m$) embeddings of the mention $m$, we apply a gated fusion mechanism inspired by prior work~\cite{wang2022multimodal, luo2023multi, song2024dual} to integrate these representations and compute the final mention embedding $\mathbf{X}_m$ as follows:
\begin{equation*}
    \mathbf{X}_m = \mathbf{g}_T \cdot \mathbf{W}_T \mathbf{T}_m^\top \;+\; \mathbf{g}_V \cdot \mathbf{W}_V \mathbf{V}_m^\top \; + \;\mathbf{W}_Z \mathbf{Z}_m^\top \in \mathbb{R}^d,
\end{equation*}
where $\mathbf{g}_T$ and $\mathbf{g}_V$ are gating coefficients that control the contributions of the textual and visual modalities, defined as:
\begin{equation*}
\mathbf{g}_T=\sigma(\mathbf{W}_T^{(\mathbf{g})} \mathbf{T}_m^\top + \mathbf{b}_T^{(\mathbf{g})})\in \mathbb{R},\;\;\;   \mathbf{g}_V=\sigma(\mathbf{W}_V^{(\mathbf{g})} \mathbf{V}_m^\top + \mathbf{b}_V^{(\mathbf{g})})\in \mathbb{R}.
\end{equation*}
Here, $\sigma(\cdot)$ is the sigmoid function.
The weight matrices $\mathbf{W}_T$, $\mathbf{W}_V$, $\mathbf{W}_Z$, $\mathbf{W}_T^{(\mathbf{g})}$, and $\mathbf{W}_V^{(\mathbf{g})}$ are modality-specific projection matrices. 
Similarly, we can compute the final embedding $\mathbf{X}_e$ of entity $e$.

\smallsection{Learning objective.}
We train \method to learn suitable representations $\mathbf{X}_m$ for mentions and $\mathbf{X}_e$ for entities using three contrastive losses.
First, we employ a mention-entity contrastive loss $\mathcal{L}_{ME}$ which aligns each mention $m$ with its corresponding ground-truth entity $e_m\in \mathcal{E}$ while separating it from unrelated entities:
\begin{equation*}
\tiny
    \mathcal{L}_{ME} = - \sum\nolimits_{m\in \mathcal{M}}\log \frac{\exp\left(\mathtt{sim}(\mathbf{X}_m, \mathbf{X}_{e_m}) / \tau_\text{cl}\right)}{\sum_{e'\in \mathcal{E}}\exp\left(\mathtt{sim}(\mathbf{X}_m, \mathbf{X}_{e'}) / \tau_\text{cl}\right)},
\end{equation*}
where $\mathcal{M}$ is the set of training mentions, $\mathtt{sim}(\cdot,\cdot)$ is the dot product similarity,
and $\tau_\text{cl}$ is a temperature term.

In addition, we introduce contrastive losses $\mathcal{L}_{MM}$ and $\mathcal{L}_{EE}$ to encourage meaningful separation of mention and entity representations in their respective embedding spaces. 
For example, $\mathcal{L}_{MM}$, which optimizes mention representations, is defined as:
\begin{equation*}
\tiny
    \mathcal{L}_{MM} = - \sum\nolimits_{m\in \mathcal{M}}\log \frac{\exp\left(\mathtt{sim}(\mathbf{X}_m, \mathbf{X}_m) / \tau_\text{cl}\right)}{\sum_{m' \in \mathcal{M}}\exp\left(\mathtt{sim}(\mathbf{X}_m, \mathbf{X}_{m'}) / \tau_\text{cl}\right)}
\end{equation*}
and $\mathcal{L}_{EE}$ for optimizing entity representations follows a similar form. 
Finally, we combine these losses into a single objective:
\begin{equation*}
    \mathcal{L} = \mathcal{L}_{ME} + \lambda_{MM} \mathcal{L}_{MM} + \lambda_{EE} \mathcal{L}_{EE}
\end{equation*}
where $\lambda_{MM}$ and $\lambda_{EE}$ are hyperparameters.

\smallsection{Candidate retrieval.}
Given the learned embeddings, we retrieve candidate entities for mention $m$ by computing the dot product between $\mathbf{X}_m$ and every entity embedding $\mathbf{X}_e$ for $e\in \mathcal{E}$.
The candidate set $\mathcal{C}(m)\subseteq\mathcal{E}$ is then the top-$K$ most similar entities, i.e.,
\begin{equation*}
    \mathcal{C}(m)={\text{Top-}K}_{e\in \mathcal{E}}\{\mathtt{sim}(\mathbf{X}_m, \mathbf{X}_e)\}.
\end{equation*}

\subsection{Stage 3. Entity Reranking}
\label{sec:entity-reranking}
After retrieving a set of candidate entities $C(m)\subseteq \mathcal{E}$ for mention $m$, we further refine the prediction by comparing the mention's triple information with that of each candidate entity.

\smallsection{Triple filtering.}
Each entity in the knowledge base can have hundreds or even tens of thousands of triples, many of which may be irrelevant or noisy for matching with the mention.
To address this, we propose a filtering scheme that retains only the most relevant triples for each candidate entity.
     Specifically, we identify the top-$n$ relations and the top-$n$ tails from the candidate entities' triples that are most similar to those of the mention, denoted as $\mathcal{R}(\mathcal{C}(m), \mathcal{T}_m)$ and $\mathcal{O}(\mathcal{C}(m), \mathcal{T}_m)$, respectively.\footnote{We compute dot product similarity between embeddings of relations (and tails) in $\mathcal{T}_m$ with those in $\mathcal{T}_e$, selecting the top-$n$ most similar ones for each element in $\mathcal{T}_m$. The filtered sets are the intersections of these top-$n$ selections.}
We then filter the triple set $\mathcal{T}_e$ of each candidate entity $e\in \mathcal{C}(m)$ as follows: 

\begin{equation*}
    {\mathcal{T}}_e^{(filt)} = \{ (e,r,o) \in \mathcal{T}_e \;|\; r\in \mathcal{R}(\mathcal{C}(m), \mathcal{T}_m) \land o \in \mathcal{O}(\mathcal{C}(m), \mathcal{T}_m) \},
\end{equation*}
ensuring that only triples with relevant relations and tails are retained in the subsequent steps.

\smallsection{Zero-shot reranking.}
Finally, we leverage an LLM to identify the best-matching entity from the candidate set $\mathcal{C}(m)$.
Specifically, we provide the mention's textual and triple-based information, along with each candidate entity's corresponding details in a step-by-step prompt: 
(1) identifying \textit{supporting triples} that serve as meaningful evidence for matching the mention with candidate entities and 
(2) determining the most appropriate entity $e_m^* \in \mathcal{C}(m)$ based on the selected supporting triples.
Formally, the final entity selection is:
\begin{equation*}
    e_m^* = \mathbf{LLM}\left( P_\text{rerank}\left(t_m,{\mathcal{T}}_m, \{t_e, \mathcal{T}_e^{(filt)}\}_{e\in \mathcal{C}(m)}\right) \right),
\end{equation*}
where $P_\text{rerank}$ is the prompt instructing the LLM to determine the most relevant entity based on both textual and triple-based information, grounding predictions in structured knowledge. For more details on the prompts, refer to Appendix \ref{app:reranking}.
Note that the filtered triples $\mathcal{T}_e^{(filt)}$ for each entity $e$ are provided, which allows the LLM to focus on the most relevant information.

\section{Experiments}
\label{sec:exp}
We conduct experiments to evaluate the performance of \method. 

\begin{table}[t]
\centering
\vspace{-2mm}
\small
\caption{
\method achieves the highest HITS@1 across three MEL datasets. 
The best results are in \textbf{bold}; the second-best are \underline{underlined}.
For results in HITS@\{3,5\} and MRR, refer to Table \ref{tab:extended-main} in Appendix~\ref{app:extended}.
}
\label{tab:baseline}
\renewcommand{\arraystretch}{0.86}
\setlength{\tabcolsep}{5.pt}
\begin{tabular}{l|ccc}
\toprule
& \textbf{WikiDiverse} &
\textbf{RichpediaMEL} &
\textbf{WikiMEL} \\ 
\midrule
CLIP~\citep{radford2021learning} & 61.21 & 67.78 & 83.23 \\
ViLT~\citep{kim2021vilt} & 34.39 & 45.85 & 72.64 \\
ALBEF~\citep{li2021align} & 60.59 & 65.17 & 78.64 \\
METER~\citep{dou2022empirical} & 53.14 & 63.96 & 72.46 \\
\midrule
DZMNED~\citep{moon2018multimodal} & 56.90 & 68.16 & 78.82 \\
JMEL~\citep{adjali2020multimodal} & 37.38 & 48.82 & 64.65 \\
VELML~\citep{zheng2022visual} & 54.56 & 67.71 & 76.62 \\
GHMFC~\citep{wang2022multimodal} & 60.27 & 72.92 & 76.55 \\
MIMIC~\citep{luo2023multi} & 63.51 & 81.02 & 87.98 \\
OT-MEL~\citep{zhang2024optimal} & 66.07 & 83.30 & \underline{88.97} \\
MELOV~\citep{sui2024melov} & 67.32 & 84.14 & 88.91 \\
M$^{3}$EL~\citep{hu2024multi} & 74.06 & 82.82 & 88.84 \\
IIER~\citep{mi2024vpmel} & 69.47 & \underline{84.63} &  88.93 \\ 
\midrule
GPT-3.5-turbo~\citep{openai2023gpt35turbo} & - & - & 73.80 \\
LLaVA-13B~\citep{liu2024visual} & - & - & 76.10 \\
GEMEL~\citep{shi2023generative} & - & - & 82.60 \\
GELR~\citep{long2024trust} & - & - & 84.80 \\
\midrule
\textbf{\ours (retrieval)} & 
\underline{82.12} \scriptsize{$\pm$ 0.21}  & 
76.40 \scriptsize{$\pm$ 0.30} & 
87.29 \scriptsize{$\pm$ 0.08} \\
\textbf{\ours (+ rerank)} & 
\textbf{88.23} \scriptsize{$\pm$ 0.29} & 
\textbf{85.21} \scriptsize{$\pm$ 0.24} & 
\textbf{90.58} \scriptsize{$\pm$ 0.25} \\
\bottomrule
\end{tabular}
\end{table}

\subsection{Experimental Settings}
\smallsection{Datasets.}
We use three public MEL datasets: WikiDiverse~\cite{wang2022wikidiverse}, RichpediaMEL~\cite{wang2022multimodal}, and WikiMEL~\cite{wang2022multimodal}.  
For a fair comparison, we adopt the dataset splits and entity set used in prior work~\cite{luo2023multi,wang2022multimodal}, selecting entities from a subset of Wikidata KB.
For each entity in these datasets, we retrieve triples from Wikidata~\footnote{\url{https://www.wikidata.org/}} using SPARQL queries.
Refer to Appendix \ref{app:dataset} for statistics and details. 

\smallsection{Evaluation metric.}
For evaluation, we measure HITS@1 unless otherwise specified. For the results on HITS@\{3,5\} and Mean Reciprocal Rank (MRR), see Table \ref{tab:extended-main} in Appendix \ref{app:extended}. 
Refer to Appendix \ref{app:eval-metric} for more details on evaluation metrics.

\smallsection{Baselines.}
We compare \ours with:
(1) retrieval-based methods that use pretrained \citep{radford2021learning,kim2021vilt,li2021align,dou2022empirical} or finetuned \citep{moon2018multimodal,adjali2020multimodal,zheng2022visual,wang2022multimodal,luo2023multi,zhang2024optimal,sui2024melov,hu2024multi,mi2024vpmel} vision-language models; and
(2) generative-based ones \citep{openai2023gpt35turbo,liu2024visual,shi2023generative,long2024trust}. 
Refer to Appendix~\ref{app:baseline} for more details. 

\vspace{4mm}
\smallsection{Implementation details.}
Following \cite{hu2024multi}, we use the pre-trained CLIP~\citep{radford2021learning} for encoding text and images, keeping it frozen during training.
We use GPT-4o-mini~\citep{openai2024gpt4omini} for generating triples and GPT-3.5-turbo~\citep{openai2023gpt35turbo} for reranking the entities.
We set $\beta = 0.5$, $\tau_\text{att} = \tau_\text{cl} = 0.1$, $\lambda_{MM} = \lambda_{EE} = 0.1$, and $K=16$ and search from ranges $p \in \{3,5\}$, and $n \in \{10,15\}$.
Refer to Appendix~\ref{app:hyperparameter} for details.

\begin{table}[t]
\centering
\vspace{-2mm}
\small  
\setlength{\tabcolsep}{2.5pt}
\caption{Ablation study on three MEL datasets demonstrating the effectiveness of each component in \method.}
\label{tab:ablation}
\begin{tabular}{lcccc}
\toprule
&  \textbf{WikiDiverse} & \textbf{RichpediaMEL} & \textbf{WikiMEL}  & $\Delta$Avg \\
\midrule
\ours (retrieval) & \textbf{82.12} & \textbf{76.40} & \textbf{87.29} & -\\
\midrule
w/o Image $\mathbf{V}$ & 81.02 &   67.19 & 80.99  &  -5.54\\
w/o Triple $\mathbf{Z}$  & 81.61  & 73.40 & \underline{85.95} &  -1.62 \\
w/o GateLayer $ \mathbf{g}_* $ & \underline{81.73} &  \underline{74.38} & 85.84& -1.29\\

\bottomrule
\end{tabular}
\end{table}

\subsection{Experimental Results}

\smallsection{Q1. Accuracy.}
\method outperforms state-of-the-art MEL methods, as shown in Table~\ref{tab:baseline}.
Notably, even at the retrieval stage, \method outperforms all baselines in WikiDiverse, and with the additional reranking stage, it achieves the best performance across all datasets, improving HITS@1 by up to 19.13\%.
These results demonstrate the effectiveness of \method's retrieve-and-rerank approach and the benefits of incorporating KG and generated triples for MEL.

\smallsection{Q2. Effectiveness.}
To assess the impact of each component in \method, we conduct ablation studies. As shown in Table~\ref{tab:ablation},
removing visual information leads to a performance drop of 5.54\% point, and excluding triple information results in a 1.62\% point decrease. 
In addition, replacing gated fusion with two linear layers degrades performance, demonstrating its importance.
Furthermore, in Table~\ref{tab:vlm}, we examine different VLMs for generating triples, where we can observe that \method's reranking scheme is effective across various VLMs, including relatively small open-source models.

\smallsection{Q3. Case studies.}
In Figure~\ref{fig:case} in Section~\ref{sec:intro}, we present a case study on how \method generates useful triples via VLM. 
In this example, \method successfully extracts relevant triples from the mention, such as \textit{<appeared\_in> - Thunderstruck} from text and \textit{<occupation> - basketball player} from the image.
These generated triples align with existing KG triples in the knowledge base, aiding in distinguishing the correct entity, which would otherwise be challenging using only textual or visual information.
This demonstrates the effectiveness of generating and leveraging triples for enhanced MEL.

\vspace{3mm}
\begin{table}[t]
\small  
\caption{Performance comparison (in terms of HITS@1; H@1) of VLMs for triple generation. 
We evaluate LLaVA-1.6-Mistral-7B, LLaVA-1.6-Vicuna-13B, and GPT-4o-mini. }
\label{tab:vlm}
\setlength{\tabcolsep}{2pt}
\begin{tabular}{cccccccc}
\toprule
\multirow{2}{*}{VLMs} & \multirow{2}{*}{stage} & \multicolumn{2}{c}{WikiDiverse} & \multicolumn{2}{c}{RichpediaMEL} & \multicolumn{2}{c}{WikiMEL} \\ 
\cmidrule{3-4}
\cmidrule{5-6}
\cmidrule{7-8}
& & H@1 &  \scriptsize{$\Delta$} & H@1 & \scriptsize{$\Delta$} & H@1  & \scriptsize{$\Delta$}\\
\midrule
\multirow{2}{*}{LLaVA-1.6-7B~\citep{liu2024llavanext}}  & retrieval & 78.99 & -& 72.89 & - & 85.09 & - \\
 & + rerank &  \underline{86.43} & +7.44  & 81.94 & +9.05 & 86.22 & +1.13 \\
 \midrule
\multirow{2}{*}{LLaVA-1.6-13B~\citep{liu2024llavanext}} & retrieval & 77.45 & - & 74.69 & - & 84.44 & -\\
 & + rerank & 85.94 & +8.49  & \underline{83.26} & +8.57 &  85.96 & +1.52 \\
 \midrule
\multirow{2}{*}{GPT-4o-mini~\cite{openai2024gpt4omini}} & retrieval & 82.12& - &  76.40   & - & \underline{87.29}  & - \\
 & + rerank & \textbf{88.23} & +6.11 &  \textbf{85.21} & +8.81 &  \textbf{90.58} & +3.29 \\ \bottomrule
\end{tabular}
\end{table}

\section{Conclusion}
\label{sec:conclusion}
In this paper, we present \method, a novel generate-retrieve-rerank framework for multimodal entity linking. 
Our analysis of KG triples in real-world knowledge bases reveals their potential for MEL. 
Based on this insight, we developed a framework that (1) generates mention triples using VLMs, (2) retrieves candidates by learning joint representations from text, image, and triples, and (3) reranks candidates based on filtered triple information, leveraging LLMs. 
Extensive experiments demonstrate that \method achieves state-of-the-art performance on three MEL benchmark datasets, validating the effectiveness of incorporating triples in MEL.

\section*{Acknowledgements}
This work was supported by Institute of Information \& Communications Technology Planning \& Evaluation (IITP) grant funded by the Korea government (MSIT) (No. RS-2024-00438638, EntireDB2AI: Foundations and Software for Comprehensive Deep Representation Learning and Prediction on Entire Relational Databases, 30\%)
(No. RS-2022-II220871, Development of AI Autonomy and Knowledge Enhancement for AI Agent Collaboration, 30\%) (No. IITP-2025-RS-2023-00253914, Artificial Intelligence Semiconductor Support Program, 30\%) (No. RS-2019-II190075, Artificial Intelligence Graduate School Program (KAIST), 10\%).

\bibliographystyle{ACM-Reference-Format}
\bibliography{ref}

\clearpage
\appendix
\section*{APPENDIX}
\label{sec:app}
\vspace{3mm}
\section{Prompt Templates}
This section provides the prompt templates used in \method for generating triples and reranking candidate entities.

\subsection{Triple Generation Prompt }
\label{app:triple-gen}
Prompt template for triple generation of mentions $P_{triple}$:

\begin{prompt}
\small{

\blue{[IMAGE]}
Given the image and text \blue{[mention sentence]}, please generate triples for the entities {list of mention word}. following the steps below:
\\
\textbf{\#\#\# Step 1: Entity Type} \\
For each entity in \blue{[list of mention words]}, identify its type, following the format:\\
- "entity\_name": type of entity\\
Type of entity can be :
person, nationality, religious group, political group, organization, country, city, state, building, airport, highway, bridge, company, agency, institution, product, event, work of art, law, language, etc.
\\
\textbf{\#\#\# Step 2: Entity Description}\\
Provide a description for each entity, following the format:\\
- "entity\_name": entity description \\
Focus on factual information that can be inferred from the image and text to describe the entity.
\\
\textbf{\#\#\# Step 3: Triples}\\
Using the type and information from steps 1 and 2, generate all possible triples for each entity in.\\
Convert the entity types and information into triples using the format, with each triple on a new line:\\
- "entity\_name" | relation1 | entity1\\
- "entity\_name" | relation2 | entity2\\
Based on the entity type and information provided in the image and text, choose the most relevant relations from the following list to generate triples:\\
"instance of", "subclass of", "part of", "has characteristic", 
"field of work", "occupation", "sex or gender", "country of citizenship", "position held", "religion or worldview", 
"member of", "owner of", "country", "capital", "continent", "located in", "industry", "participant", "genre", "named after"

}
\end{prompt}

The triple generation prompt guides the VLM through a step-by-step process. First, for entity type identification, we reference the NER (Named Entity Recognition) categories from \citep{weischedel2013ontonotes}. This identified type is further incorporated into a triple with the relation "instance of." Next, rather than generating triples directly, we use entity descriptions as a reasoning step. To better align with existing KG triples $\mathcal{T}_e$, we select 20 relation types based on their frequency within $\mathcal{T}_e$ and their semantic relevance to mention contexts, providing them as references. We generate triples for each mention sentence, processing all the mention words in the sentence one at a time. This approach considers the contextual relationships between mentions and is also efficient.

\vspace{3mm}
\subsection{Reranking Prompt}
\label{app:reranking}
Prompt template for zero-shot reranking $P_{rerank}$ :  

\begin{prompt}
\small{
Given the context below, please identify the most corresponding entity from the list of candidates.

\textbf{Context:} \blue{[mention sentence]}

\textbf{Candidate Entities:}\\
\blue{[$K^{th}$ entity name]} ( \blue{[QID]} ) : \blue{[description]}\\
- Triple: \blue{\{list of KG triples\}} \\ 
. . . \\\
\blue{[$1^{st}$ entity name]} ( \blue{[QID]} ) : \blue{[description]}\\
- Triple: \blue{[list of KG triples]} \\ 
\\
\textbf{Context:} \blue{[mention sentence]}

\textbf{Target Entity:} \blue{[mention words]}: \blue{[generated description]}\\
- Triple: \blue{[list of generated triples]} \\ 

First, read the context and the target entity. Then, review the candidate entities and their descriptions.\\
From the candidate entities, select the \textit{supporting triples} that align with the context and the target entity. (Note that triples may contain inconsistent information.)\\
Based on the selected supporting triples, identify the most relevant entity that best matches the given sentence context.}
\end{prompt}

The entity reranking prompt instructs LLMs to identify \textit{supporting triples} from filtered entity triples $\mathcal{T}_e^{(filt)}$ and filtered mention triples $\mathcal{T}_m^{(filt)}$  to determine the most relevant entity match from candidates $\mathcal{C}(m)$. The candidates are presented following the order from the candidate retrieval stage, but in reverse order, which we empirically found to improve performance. The prompt also leverages entity description and the generated mention descriptions from step 2 of the triple generation process as additional context.

\section{Experimental Setup}
This section presents the experimental setup, including the evaluation metrics for baselines, hyperparameter configurations for each dataset, and additional experimental results

\subsection{Evaluation Metrics}
\label{app:eval-metric}
We evaluate \method using HITS@k and MRR, defined as
\begin{align}  
HITS@k &= \frac{1}{N}\sum_{i}I(rank(i) \leq k), \\  
MRR &= \frac{1}{N}\sum_{i}\frac{1}{rank(i)},  
\end{align}  
where $N$ is the total number of test instances, $rank(i)$ denotes the rank of the correct entity for the $i$-th instance, and $I(\cdot)$ is an indicator function. 

\subsection{Baselines}
\label{app:baseline}
We compare the performance of \ours with several baseline methods, which are grouped into two categories:  \\
\textbf{Retrieval-based Methods:}
\begin{itemize}[left=0em]
    \item \textbf{CLIP}~\citep{radford2021learning} aligns visual and textual inputs using two transformer-based encoders trained on extensive image-text pairs with a contrastive loss.
    \item \textbf{ViLT}~\citep{kim2021vilt} employs shallow embeddings for text and images, emphasizing deep modality interactions through transformer layers.
    \item \textbf{ALBEF}~\citep{li2021align} integrates visual and textual features via a multimodal transformer encoder, utilizing image-text contrastive loss and momentum distillation for improved learning from noisy data.
    \item \textbf{METER}~\citep{dou2022empirical} explores semantic relationships between modalities using a co-attention mechanism comprising self-attention, cross-attention, and feed-forward networks.
    \item \textbf{DZMNED}~\citep{moon2018multimodal} is the first method for MEL, integrates visual features with word-level and character-level textual features using an attention mechanism.
    \item \textbf{JMEL}~\citep{adjali2020multimodal} extracts and fuses unigram and bigram textual embeddings, jointly learning mention and entity representations from both textual and visual contexts.
    \item \textbf{VELML}~\citep{zheng2022visual} utilizes VGG-16 for object-level visual features and a pre-trained BERT for text, combining them via an attention mechanism.
    \item \textbf{GHMFC}~\citep{wang2022multimodal} employs hierarchical cross-attention to capture fine-grained correlations between text and images, optimized through contrastive learning.
    \item \textbf{MIMIC}~\citep{luo2023multi} proposes a multi-grained multimodal interaction network that captures both global and local features from text and images, enhancing entity disambiguation through comprehensive intra- and inter-modal interactions.
    \item \textbf{OT-MEL}~\citep{zhang2024optimal} addresses multimodal fusion and fine-grained matching by formulating correlation assignments between multimodal features and mentions as an optimal transport problem, with knowledge distillation.
    \item \textbf{MELOV}~\citep{sui2024melov} optimizes visual features in a latent space by combining inter-modality and intra-modality enhancements, improving consistency between mentions and entities.
    \item \textbf{M$^3$EL}~\citep{hu2024multi} introduces a multi-level matching network for multimodal feature extraction, intra-modal matching, and bidirectional cross-modal matching, enabling comprehensive interactions within and between modalities.

    \item \textbf{IIER}~\citep{mi2024vpmel} improves entity linking by using visual prompts as guiding texture features to focus on specific local image regions, and by generating auxiliary textual cues using a pre-trained Vision-language model (VLM).

\end{itemize}

\noindent\textbf{Generative-based Methods:}
    \begin{itemize}[left=0em]
    \item \textbf{GPT-3.5-turbo}~\citep{openai2023gpt35turbo} is a large language model (LLM), and we utilize the results reported by GEMEL.  
    \item \textbf{LLaVA-13B}~\citep{liu2024visual} is a vision-language model (VLM), and we utilize the results reported by GELR.
    \item \textbf{GEMEL}~\citep{shi2023generative} leverages large language model (LLM) to directly generate target entity names, aligning visual features with textual embeddings through a feature mapper.
    \item \textbf{GELR}~\citep{long2024trust} enhances the generation process by incorporating knowledge retriever, improving accuracy through the retrieval of relevant context from a knowledge base.
    \end{itemize}

\newpage
\subsection{Hyperparameter Settings}
\label{app:hyperparameter}

Table~\ref{tab:hyper} shows the hyperparameter settings used for each dataset in our experiments.

\begin{table}[h]
\centering
\small
\caption{Hyperparameter settings.}
\begin{tabular}{c|ccc}
\toprule
\textbf{Hyperparameter} & \textbf{WikiDiverse} & \textbf{RichpediaMEL}  & \textbf{WikiMEL}\\
\midrule
$\beta$ & 0.5 & 0.5 & 0.5 \\
$\tau_\text{att}$, $\tau_\text{cl}$  & 0.1 & 0.1 & 0.1 \\
$\lambda_{MM}$, $\lambda_{EE}$  & 0.1 & 0.1 & 0.1 \\
$K$ & 16 & 16 & 16 \\
$p$ & 3 & 3 & 5 \\
$n$ & 15 & 10 & 15 \\
\bottomrule
\end{tabular}
\label{tab:hyper}
\end{table}

\section{Dataset Statistics}
\label{app:dataset}
We evaluate \method on three MEL datasets: WikiDiverse~\citep{wang2022wikidiverse}, RichpediaMEL~\citep{wang2022multimodal} and WikiMEL~\citep{wang2022multimodal}. We use a subset of Wikidata as KB, following \citep{wang2022multimodal}, and retrieve KG triples via SPARQL queries. Table \ref{tab:data-stat} summarizes dataset statistics and retrieved KG triples.

\begin{table}[h]
\centering
\small
\caption{Statistics of three MEL datasets.}
\setlength{\tabcolsep}{5pt} 
\begin{tabular}{c|cccc}
\toprule
\textbf{Datasets} & \textbf{WikiDiverse} & \textbf{RichpeidaMEL} & \textbf{WikiMEL} \\
\midrule
\# sentences & 7,405 & 17,724 & 22,070 \\
\# mentions & 15,093 & 17,805 & 25,846 \\
\# KG triples & 60,842,321 & 32,761,864 & 65,131,860 \\
\# candidate entities$^\dagger$ & 132,460 & 160,935& 109,976  \\
\# total entities$^\ddagger$ & 776,407 &  831,737 &  761,343 \\
\# relation & 1,322 & 1,288 & 1,289 \\
\bottomrule
\end{tabular}
\label{tab:data-stat}
\vspace{1mm}
\raggedright \\
\footnotesize{$^\dagger$ All entities in a subset of Wikidata KB used as candidates.} \\
\footnotesize{$^\ddagger$ Includes candidate entities and all tail entities from retrieved KG triples.}
\vspace{5mm}
\end{table}

\section{Extended Experimental Results}
\label{app:extended}
Table~\ref{tab:extended-main} shows extended experimental results, including HITS@\{1,3,5\} and MRR. The results are reported as the mean $\pm$ standard deviation across three experimental runs. For the reranking results, the top-1 retrieved entity is replaced by the entity selected during the reranking stage, while the ranking order of the remaining candidates is preserved. We obtained the baseline results from \citep{luo2023multi}, while the results for \citep{zhang2024optimal,sui2024melov,hu2024multi,mi2024vpmel} were sourced from the original papers. Results for \citep{shi2023generative,long2024trust} were also taken from the original papers, along with those for \citep{openai2023gpt35turbo,liu2024visual}.

\begin{table*}[t]
    \centering
    \small  
    \caption{Evaluation results on three MEL datasets. H@k denotes HITS@k, MRR denotes Mean Reciprocal Rank. The best results are in \textbf{bold}; the second-best are \underline{underlined}. }
\setlength{\tabcolsep}{2.3pt}
\begin{tabular}{l|cccc|cccc|cccc}
    \toprule
    &  
    \multicolumn{4}{c|}{\textbf{WikiDiverse}} &
    \multicolumn{4}{c|}{\textbf{RichpediaMEL}} & 
    \multicolumn{4}{c}{\textbf{WikiMEL}}  \\
    \cmidrule{2-5}
    \cmidrule{6-9}
    \cmidrule{10-13}
    ~ & H@1 & H@3 & H@5 & MRR & H@1 & H@3 & H@5 & MRR & H@1 & H@3 & H@5 & MRR \\
    \midrule
    CLIP~\citep{radford2021learning} & 61.21 & 79.63 & 85.18 & 71.69 & 67.78 & 85.22 & 90.04 & 77.57 & 83.23 & 92.10 & 94.51 & 88.23 \\ 
    ViLT~\citep{kim2021vilt}  & 34.39 & 51.07 & 57.83 & 45.22 & 45.85 & 62.96 & 69.80 & 56.63 & 72.64 & 84.51 & 87.86 & 79.46 \\
    ALBEF~\citep{li2021align} & 60.59 & 75.59 & 81.30 & 69.93 & 65.17 & 82.84 & 88.28 & 75.29 & 78.64 & 88.93 & 91.75 & 84.56 \\
    METER~\citep{dou2022empirical} & 53.14 & 70.93 & 77.59 & 63.71 & 63.96 & 82.24 & 87.08 & 74.15 & 72.46 & 84.41 & 88.17 & 79.49 \\
    \midrule
    DZMNED~\citep{moon2018multimodal} & 56.90 & 75.34 & 81.41 & 67.59 & 68.16 & 82.94 & 87.33 & 76.63 & 78.82 & 90.02 & 92.62 & 84.97 \\ 
    JMEL~\citep{adjali2020multimodal}  & 37.38 & 54.23 & 61.00 & 48.19 & 48.82 & 66.77 & 73.99 & 60.06 & 64.65 & 79.99 & 84.34 & 73.39 \\ 
    VELML~\citep{zheng2022visual} & 54.56 & 74.43 & 81.15 & 66.13 & 67.71 & 84.57 & 89.17 & 77.19 & 76.62 & 88.75 & 91.96 & 83.42 \\ 
    GHMFC~\citep{wang2022multimodal} & 60.27 & 79.40 & 84.74 & 70.99 & 72.92 & 86.85 & 90.60 & 80.76 & 76.55 & 88.40 & 92.01 & 83.36 \\
    MIMIC~\citep{luo2023multi} & 63.51 & 81.04 & 86.43 & 73.44 & 81.02 & 91.77 & 94.38 & 86.95 & 87.98 & 95.07 & 96.37 & 91.82 \\
    OT-MEL~\citep{zhang2024optimal} & 66.07 & 82.82 & 87.39 & 75.43 & 83.30 & 92.39 & 94.83 & \underline{88.27} & \underline{88.97} & \underline{95.63} & \textbf{96.96} & \underline{92.59} \\
    MELOV~\citep{sui2024melov} & 67.32 & 83.69 & 87.54 & 76.57 & 84.14 & \underline{92.81} & 94.89 &  \textbf{88.80} & 88.91 &  95.61 &  96.58 & 92.32 \\
    M$^{3}$EL~\citep{hu2024multi} & 74.06 & 86.57 & 90.04 & 81.29  & 82.82 &  92.73 & \textbf{95.34} 
    & 88.26  & 88.84 & 95.20 & 96.71 & 92.30  \\
    IIER ~\citep{mi2024vpmel} & 69.47 & 84.43  & 88.79  & -& \underline{84.63} & \textbf{93.27} & \underline{95.30} & - & 88.93 & \textbf{95.69} & \underline{96.73} & - \\ 
    \midrule
GPT-3.5-turbo~\citep{openai2023gpt35turbo} & - & - & - & - & - & - & -   & - & 73.80 & - & - & - \\
LLaVA-13B~\citep{liu2024visual}  & - & - & - & - & - & - & - &  - & 76.10& - & - & - \\
GEMEL~\citep{shi2023generative}$^{\dagger}$  & - & - & - & - & - & - & -  &  - & 82.60   & - & - & - \\
GELR~\citep{long2024trust}$^{\dagger}$  & - & - & - & - & - & - & - &  - & 84.80 & - & - & - \\
\midrule
\textbf{\ours(retrieval)} & 
\underline{82.12}\scriptsize{$\pm$0.21} & \underline{90.28}\scriptsize{$\pm$0.17} & \underline{92.07}\scriptsize{$\pm$0.05} & \underline{86.00}\scriptsize{$\pm$0.16} & 
76.40\scriptsize{$\pm$0.30} & 85.92\scriptsize{$\pm$0.28} & 88.82\scriptsize{$\pm$0.15} & 80.94\scriptsize{$\pm$0.45} & 
87.29\scriptsize{$\pm$0.08} & 92.47\scriptsize{$\pm$0.34} & 93.94\scriptsize{$\pm$0.27} & 89.99\scriptsize{$\pm$0.25} \\
\textbf{\ours(+rerank)} & 
\textbf{88.23}\scriptsize{$\pm$0.29} & \textbf{92.82}\scriptsize{$\pm$0.06} & \textbf{93.61}\scriptsize{$\pm$0.17} & \textbf{90.84}\scriptsize{$\pm$0.13} & 
\textbf{85.21}\scriptsize{$\pm$0.24} & 89.85\scriptsize{$\pm$0.20} & 91.32\scriptsize{$\pm$0.19} & 88.08\scriptsize{$\pm$0.21} & 
\textbf{90.58}\scriptsize{$\pm$0.25} & 95.18\scriptsize{$\pm$0.29} & 95.87\scriptsize{$\pm$0.24} & \textbf{93.04}\scriptsize{$\pm$0.26} \\
    \bottomrule
    \multicolumn{4}{l}{$^{\dagger}$: Those that fine-tune LLMs.} \\
    \end{tabular}
    \label{tab:extended-main}
\end{table*}

\end{document}